\documentclass{vietnam}

\def\be{\begin{equation}}
\def\ee{\end{equation}}
\def\bea{\begin{eqnarray}}
\def\eea{\end{eqnarray}}

\def\hh95{$h_{95}$}
\def\hh125{$h_{125}$}
\def\HH320{$H_{320}$}
\def\HH650{$H_{650}$}

\usepackage[capitalise]{cleveref}
\creflabelformat{equation}{#2\textup{#1}#3}

\newcommand{\Photo}{}

\newcommand{\lsim}{\raisebox{-0.13cm}{~\shortstack{$<$ \\[-0.07cm]
      $\sim$}}~}
\newcommand{\bmF}{\mathbf{F}}

\begin{document}
\title{Resonances all over the place?\\
{\sl To appear in the proceedings of the 29th International Symposium on Particles, Strings and Cosmology (PASCOS 2024) in Quy Nhon, Vietnam, 7-13 July, 2024.}}

\author{ Anirban Kundu \footnote{Electronic address: akphy@caluniv.ac.in} }

\address{Department of Physics, University of Calcutta,
92 Acharya Prafulla Chandra Road, Kolkata 700009, India}

\author{ Poulami Mondal \footnote{Electronic address: poulami.mondal1994@gmail.com}}

\address{Department of Physics, Indian Institute of Technology Kanpur, 
Kanpur 208016, India}

\author{ Gilbert Moultaka \footnote{speaker} \footnote{Electronic address: gilbert.moultaka@umontpellier.fr}}

\address{Laboratoire Univers \& Particules de Montpellier (LUPM), 
Universit\'e de Montpellier, CNRS, Montpellier, France}

\maketitle\abstracts{
We provide a possible interpretation of excesses reported by ATLAS and CMS at around 95GeV, 650GeV and possibly 320GeV, in terms of CP-even scalars. In particular, the combined {\sl global} statistical significances of independent indications for a 650GeV object reach the $4\sigma$ level! While this seems sufficient incentive for a further investigation, this object cannot be fitted in tradional singlet or doublet extensions of the Standard Model. It requires by itself a larger extension with doubly-charged scalars, that naturally fits the two other excesses on top of the SM-like 125~GeV Higgs. We describe the minimal model and give some numerical illustrations.}

\section{Introduction}
The Higgs-like boson discovery at the LHC \cite{ATLAS:2012yve,CMS:2012qbp} 
occured alongside another major event: the non-discovery of phenomena beyond the Standard Model at the TeV scale motivated by seemingly strong theoretical arguments such as  {\sl naturalness} and/or the {\sl hierarchy problem} of physical scales. 
While this could mean a number of different things the discussion of which lies outside the present note, the search for new resonances by ATLAS and CMS is a precious tool that might perhaps shed some light on the right way to go in the absence of a clear theoretical guide, and given the several other shortcomings of the Standard Model (SM).  

Hereafter we report on a possible interpretation of the indications for excesses around 95~GeV ($h_{95}$), 650~GeV ($H_{650}$) and indirectly 350~GeV ($H_{320}$), in terms of electrically neutral CP-even scalar states together with the 125~GeV Higgs ($h_{125}$). See reference \cite{Kundu:2024} for details.

\section{The Indications}
\label{sec:indications}

To assess the various indications we will use the significances quoted officially by the experiments and adopt the conservative attitude of combining only the global ones when available. For this we combine the p-values corresponding to the global significances of the indipendent channels for a given resonance, using a standard combiner, e.g. Fisher's method \cite{Fisher:1934}. \Cref{tab:significance} recalls the various processes for each resonance, the local and global significances as quoted by the referenced experimental papers or notes, as well as our combined {\sl global} significances. 
It is noteworthy that $H_{650}$, which attracted much less attention than $h_{95}$ in the recent years, has a combined global significance higher than that for $h_{95}$. Even though much lower than what one obtains when neglecting the look-elsewhere effect~\cite{Biekotter:2022jyr,Kundu:2022bpy}, we see the level of these significances a sufficient motivation to consider all the hints simultaneously, in particular given the $4\sigma$ for the $H_{650}$ resonance.  Indeed, if the latter is interpreted as an elementary scalar, its reported properties \cite{CMS:2022bcb} would require the presence of two extra
CP-even scalars on top of $h_{125}$, as well as the expectation of CP-odd scalars, if it is to be embeded in a consistent theoretical framework as we will argue in \cref{sec:models}. It is in that sense that the indications for $h_{95}$, $H_{320}$ (and $A_{400}$) are welcome even though with somewhat low significances.

\begin{table}[h]
\caption{The existing indications for some scalar resonances, apart from $h_{125}$.
}
\begin{center}
\resizebox{.8\textwidth}{!}{
\begin{tabular}{||c||c||c|c|c||c||}
\hline
New scalar & Process studied & Local & Global & Combined & Reference \\
      &    & Significance & Significance & Significance & \\
\hline
 $h_{95}$     & $\to \gamma\gamma$ & $2.9\, \sigma$ & $1.3\, \sigma$ &            & \cite{CMS:2018cyk,CMS:2015ocq}\\
                    & $\to \tau^+\tau^-$  & $2.6$--$3.1\, \sigma$& $2.3$--$2.7\, \sigma$ &     $2.4$--$2.75\, \sigma$          &   \cite{CMS:2022rbd,CMS:2022goy} \\
                    & $Z^*\to Zh_{95}\to Zb\bar{b}$ & $2.3\, \sigma$    & not quoted  &  $3.1$--$3.4\, \sigma$   &       
                    \cite{LEPWorkingGroupforHiggsbosonsearches:2003ing}\\
\hline
$H_{650}$   & VBF\,, $\to W^+W^-$  & $3.8\, \sigma$ &  $2.6\pm 0.2\, \sigma$  &            &  \cite{CMS:2022bcb} \\
                   & $\to ZZ$  &  $2.4\, \sigma$   &  $0.9\, \sigma$  &  $4.08^{+0.12}_{-0.11} \, \sigma$        &   \cite{ATLAS:2020tlo,ATLAS:2021kog} \\
                   & $\to h_{95}h_{125}$  &  $3.8\, \sigma$      & $2.8\, \sigma$  &       &     \cite{CMS:2022tgk}        \\
                   & $\to A_{400}Z \to \ell^+\ell^-t\bar{t}$ & $2.85\sigma$ & $2.35\sigma$ &  & \cite{ATLAS:2023zkt}\\
                   \hline
$A_{400}$   & $\to t\bar{t}$   &    $3.5\, \sigma$    & $1.9\, \sigma$ &     $3.17\, \sigma$    & \cite{CMS:2019pzc}    \\
                   & $\to ZH_{320}\to Zh_{125}h_{125}$ & $3.8\, \sigma$ & $2.8\, \sigma$   &   & \cite{ATLAS:2022wti}  \\
\hline
\end{tabular}}
\label{tab:significance}
\end{center}
\end{table}


For later use, we recall here the reported signal strengths for $h_{95}$: 
\begin{equation}
\!\!\!\!\!\!\!\!\!\!\!\!\!\!\!\!\!
\mu_{\gamma\gamma} = \frac{\sigma(pp \to h_{95} \to \gamma\gamma)}{\sigma(pp\to \phi\to
\gamma\gamma)} = 0.33^{+0.19}_{-0.12}\, \ \ \ \ \ \ \ \   [{{\rm LHC(CMS)} } ]
\label{eq:CMSh95gg}
\end{equation}
\begin{equation}
\!\!\!\!\!\!\!\!\!\!\!\!\!\!\!\!\mu_{\tau\tau} = \frac{\sigma(pp \to h_{95} \to\tau^+\tau^-)}{\sigma(pp\to \phi \to
\tau^+\tau^-)} = 1.2\pm 0.5\,,  \ \ \ \  [{{\rm LHC(CMS)} } ]\label{eq:CMSh95tautau}
\end{equation}
\begin{equation}
\!\!\!\!\!\!\!\!\!\!\!\!\!\!\!\!\mu_{b\bar{b}} = \frac{\sigma(e^+e^- \to Zh_{95} \to Zb\bar{b})}{\sigma(e^+e^-\to Z\phi \to Zb\bar{b})} =
0.117\pm 0.057\,, \ \ \ \ \, [{{\rm LEP} } ]\label{eq:LEPh95}
\end{equation}
The relevance of the $95$~GeV indication at LEP has been recently disputed \cite{Janot:2024ryq}. We will take this into account by comparing the results of our study with and without \cref{eq:LEPh95}. In \cref{tab:significance} we give the combined significances for both cases, (using though the local significance quoted by the LEP working group \cite{LEPWorkingGroupforHiggsbosonsearches:2003ing} where the global one was not provided).

\begin{figure}[h]
\begin{center}
{\includegraphics[width=.4\linewidth]{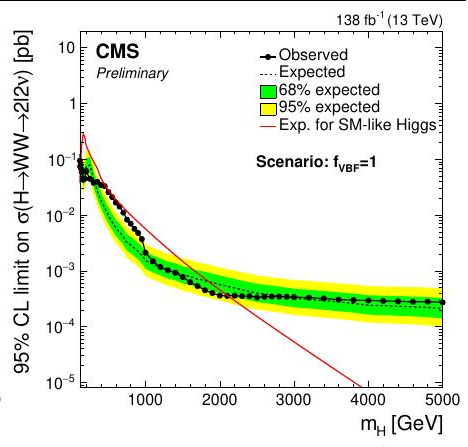}}
\end{center}
\caption{Extracted from Figure 4 of the CMS note{~}\protect\cite{CMS:2022bcb}}
\label{fig:CMS}
\end{figure}

As for $H_{650}$, the excess found by CMS \cite{CMS:2022bcb} for $ \sigma(p p \to W^+ W^- \to 2\ell2\nu)$ is reported to be consistent with the production and decay of a heavy SM-like scalar through  $p p \to H_{650} \to W^+W^-)$, where the data and the SM-like Higgs expectation intersect around a mass of 650~GeV with a $\sigma(p p \to H_{650} \to W^+W^-)$  cross-section around 160~fb, {\sl but} under the assumption that the latter  proceeds only through VBF, see \cref{fig:CMS}. Rather than using this cross-section
that relies on VBF and ggF experimental categorizations to which we do not have access, one can proceed differently by exploiting the abovementioned intersection to equate the SM-like and non-SM cross-sections at around 650~GeV independently from their actual magnitudes:  
\begin{equation}
 \sigma_{\rm VBF}\times {\rm BR}_{H_{650} \to W^+W^-} = c \sigma^{(\rm SM)}_{\rm VBF}\times {\rm BR^*}^{(\rm SM)}_{H_{650} \to W^+W^-}. \label{eq:xsec-data-SM}
\end{equation}
Here the starred BR on the right-hand side indicates evaluation without the $t\bar t$ contribution to comply with the CMS scenario of only-VBF-production signal, and $c$ is an order 1 factor that can account for uncertainties.  
This equation leads to a strong correlation among the (reduced) couplings of $H_{650}$ to $ZZ$, $WW$ and $t\bar t$ appearing on the right-hand side of \cref{eq:xsec-data-SM}, provided a split of $\sigma^{(\rm SM)}_{\rm VBF}$ into separate $WW$- and $ZZ$-fusion  contributions is performed through VBF-cuts that minimize intereference contributions.  In fact, consistency leads to a narrow range for the coupling to $WW$ \cite{Kundu:2024} :
\begin{equation}
0.96 \sqrt{c} \, gM_W \lsim \left| g_{{}_{WWH_{650}}} \right| \lsim  (0.05 + 0.95 \sqrt{c})\,gM_W  , \label{eq:WWH650min}
\end{equation}
where $g$ is the weak gauge coupling.

As for $H_{320}$ and $A_{400}$, we take them only as a qualitative indication, the first completing the set of four CP-even states, the second as one of the two CP-odd states,  needed for a coherent model description, as we argue in the next section.  Information about the 
$H_{320} Z Z$ coupling from the process shown in the last line of \cref{tab:significance}
is somewhat screened by the theoretical uncertainty on the magnitude of a supposed $H_{320} h_{125} h_{125}$ coupling that would originate from an extended scalar potential with more freedom than in the case of couplings to gauge bosons. Couplings to $ZZ$ and $WW$ should however remain sufficiently smaller than $H_{320} h_{125} h_{125}$ to account for the non-observation of decays to gauge bosons \cite{ATLAS:2022wti}. As a guide for the scans shown in \cref{sec:pos-sol}, we will require the reduced couplings to $W$'s and $Z$'s to satisfy
\begin{equation}
|\kappa_W^{H_{320}}|, |\kappa_Z^{H_{320}}| \lsim .45 \, . \label{eq:H320-constraints}
\end{equation} 
Let us also recall the values of the fitted coupling modifiers for the SM-like $h_{125}$ resonance. We take for definitness \cite{CMSnature}:
\begin{equation}
\kappa_W^{h_{125}} = 1.02\pm 0.08\,,\ \ \kappa_Z^{h_{125}} = 1.04\pm 0.07\,,\ \ 
\kappa_t^{h_{125}} = 1.01^{+0.11}_{-0.10}\,,\ \ \kappa_b^{h_{125}} = 0.99^{+0.17}_{-0.16}\,. \label{eq:CMS-SMkappas}
\end{equation}
We will refer to the list of requirements \cref{eq:CMSh95gg,eq:CMSh95tautau,eq:WWH650min,eq:H320-constraints,eq:CMS-SMkappas} collectively as `the constraints', and refer seperately to the controversial requirement \cref{eq:LEPh95} as `LEPh95', and to the CMS indication \cref{eq:xsec-data-SM}. 

\section{Models}
\label{sec:models}

\subsection{A Sum Rule}

As well known, massive gauge boson scattering requires the presence of scalar particles in order to tame the high-energy behaviour of the corresponding cross-sections that potentially violate unitarity. This leads in a renormalizable theory with an extended scalar sector beyond the SM to constraints among the scalar--gauge-boson--gauge-boson couplings \cite{Gunion:1990kf}. In particular one has the general sum rule from the unitarization of the longitudinal $W$ scattering $W_L W_L \to W_L W_L$ at the tree-level: 
\begin{equation}
 {g^{{(SM)} 2}_{{}_{WWh}}} = \sum_i g_{{}_{W^+ W^- \phi^0_i}}^2 
- \sum_k |g_{{}_{W^- W^- {\phi_k}^{++}}}|^2,  \label{eq:sumrule}
\end{equation}
where $g^{(SM)}_{{}_{WWh}}$ is the SM $WWh$ coupling and we assumed CP conservation and all the neutral scalars $\phi^0_i$ CP-even, and for simplicity $\rho=1$ at the tree-level. Since the sum in the first term on the right-hand side of \cref{eq:sumrule} must include a SM-like Higgs with a coupling very close to 
$g^{(SM)}_{{}_{WWh}}$, the presence of an extra coupling in the sum as
large as $g_{{}_{WWH_{650}}}$, cf. \cref{eq:WWH650min}, necessitates the compensating effect of at least one doubly-charged state as to satisfy the above sum rule. Assuming all these scalars to be elementary particles, the above consequence becomes unavoidable. It then rules out any scalar extension of the SM by only (i)  $\rm SU(2)_L$ singlets or (ii) $\rm SU(2)_L$ doublets, since doubly-charged states cannot be included consistently, but also other extensions as we will see hereafter.  

\subsection{Just One or Two Triplets Would not Work}
\label{sec:notriplets}
The minimal way of including a doubly-charged scalar respecting the gauge symmetries of the model would be by adding an $\rm SU(2)_L$ complex triplet to the Higgs sector of the SM. However, with just one such a multiplet like in the Type-II Seesaw Model, not only there would be just one CP-even scalar on top of the SM-like state, say the $H_{650}$ or $h_{95}$, but it would also be too weakly coupled to fermions and to pairs of gauge bosons to be singly produced at the LHC---a direct consequence of a tiny triplet VEV that ensures the $\rho \simeq 1$ requirement---This is in contradiction with the previously discussed indications  even if one would give up on the other resonances.

Adding to the SM one complex and one real $\rm SU(2)_L$ triplets allows larger triplet VEVs consistent with $\rho \simeq 1$ provided the two VEVs are taken equal as in the 
Georgi-Machacek model (GM). However, now the coupling to pairs of gauge bosons are fixed by a residual
custodial symmetry of the scalars kinetic terms. The CP-even $H_{650}$ will be either a custodial singlet or part of the custodial five-plet (more about this in the next section).
This would correspond to
$g_{{}_{ZZH_{650}}}/g_{{}_{WWH_{650}}} \simeq 1.3$ or $2.6$ respectively, contradicting the experimental indications \cite{CMS:2022bcb,ATLAS:2020tlo,ATLAS:2021kog} that favor a reversed hierarchy between the two couplings. Moreover, such a setup would allow for only 3 CP-even scalars.

\subsection{The extended GM}

We see from the above arguments that fitting the properties of $H_{650}$ drives us naturally to further extensions which will allow to embed all four scalar states, a bonus rather than a requirement! Adding one gauge singlet would not solve the mismatch noted at the end of the previous section. Let us try an extra doublet $\Phi_2$, on top of the SM one
$\Phi_1$ and the two triplets $\chi$ and $\xi$: 
\begin{eqnarray}
&\Phi_1 = 
\left(\begin{array}{cc}
\phi^{0 *}_1 & \phi^+_1  \\
- \phi^{+ *}_1 & \phi^0_1 \\
\end{array}
 \right), 
\Phi_2 = 
\left(\begin{array}{cc}
\phi^{0 *}_2 & \phi^+_2  \\
- \phi^{+ *}_2 & \phi^0_2 \\
\end{array}
\right), 
 X = 
{\left(
\begin{array}{ccc}
\chi^{0 *} & \xi^+ & \chi^{++}\\
-\chi^{+ *} & \xi^0& \chi^+\\
\chi^{++ *} & -\xi^{+ *}& \chi^0\\
\end{array}
\right) }. \label{eq:multiplets}
\end{eqnarray}
We have reexpressed each of the two doublets as bi-doublets and collected the two triplets in a single bi-triplet---including consistently the complex conjugate representations of the doublets and the complex triplet. One has 17 real scalar degrees of freedom, 3 of which are taken by the weak gauge boson longitudinal components, and the remaining 14 combine into physical states (4 CP-even, 2 CP-odd, 6 charged and 2 doubly-charged states). The matrix forms in Eq.~\ref{eq:multiplets} set the stage for an extension of the (approximate) custodial symmetry present in the SM to the present model: An $\rm SU(2)_L \times \rm SU(2)_R$ symmetry transformation $\Phi_i \to U_L^{(2)} \Phi_i U_R^{(2)}$ and $X \to U_L^{(3)} X U_R^{(3)}$~\footnote{where $U^{(2),(3)}_{L, R}$ denote elements of $\rm SU(2)_L$ and a new global $\rm SU(2)_R$
group respectively in the fundamental and vector representations of $SU(2)$}, keeps invariant the $\rm SU(2)_L$-gauge kinetic terms of the scalars. It is, however,  spontaneously broken, after electroweak-symmetry breaking, to a residual $\rm SU(2)_{custodial}$ corresponding to the restricted above transformations with 
$U^{(2),(3)}_L=U^{(2),(3)}_R$ that keep the vacuum invariant provided that the VEVs of the two triplets are equal. The $\rm SU(2)_{custodial}$ transformation being the simultaneous product of  $\rm SU(2)_L$ and $\rm SU(2)_R$ one can rearrange the doublets and triplets components of Eq.~\ref{eq:multiplets} in irreducible representations of  the products $2 \otimes 2$ and $3 \otimes 3$ respectively, that is singlets and triplets, and singlets, triplets and five-plets of the custodial group. Here we are only intersted in the neutral CP-even states that are necessarily either custodial singlets, corresponding to the vacuum shifted fields since the vacuum is invariant, or a member of the custodial five-plet
with zero VEV---the CP-odd scalars are in custodial triplets akin to the Goldstone bosons and the fact that the three $\rm SU(2)_L$ gauge bosons collected in the vector $\mathbf{W}$ form themselves an $\rm SU(2)_{custodial}$  triplet. The invariance of the kinetic terms after electroweak-symmetry breaking noted above fixes uniquely the gauge-gauge-scalar couplings by writing down custodial scalar terms with proper normalizations \cite{KMP22,Kundu:2024}:
\begin{equation}
{\cal L}_{\rm kin} \supset \frac{g^2v_1}{2\sqrt{2}} {\rm Re} \phi_1^0 \, \mathbf{W}\cdot \mathbf{W}
+ \frac{g^2v_2}{2\sqrt{2}} {\rm Re} \phi_2^0 \, \mathbf{W}\cdot \mathbf{W} 
+ \frac{2g^2 u}{\sqrt{3}} S_3\, \mathbf{W}\cdot \mathbf{W} 
+ g^2 u \left[ \mathbf{W}\otimes \mathbf{W}\right]\cdot \bmF\, \label{eq:Lkin}
\end{equation}
where $g$ is the $\rm SU(2)_L$ gauge coupling, the $\phi_i^0$'s denote the fields after the vacuum shifts $v_1, v_2$, $S_3$ a combination of $\xi^0$ and ${\rm Re} \chi^0$ after a common vacuum shift $u$, and $\bmF$ a custodial five-plet whose neutral component $F_0$ is a combination of $\xi^0$ and ${\rm Re} \chi^0$ with zero VEV; $\mathbf{W}\cdot \mathbf{W}$ and $\mathbf{W}\otimes \mathbf{W}$ denote respectively the custodial singlet and five-plet formed from $\mathbf{W}$. \Cref{eq:Lkin} gives the couplings to $W^+W^-$, but also to $ZZ$ substituting the neutral component of $\mathbf{W}$ by $Z/\!\cos \theta_W$, $\theta_W$ being the weak mixing angle. It follows that if any of the scalar states in  \cref{eq:Lkin} coincides with the physical $H_{650}$, then we are back to the problem noted at the end of \cref{sec:notriplets}. This would typically happen if the scalar potential were invariant under the $\rm SU(2)_L \times \rm SU(2)_R$ as in the conventional GM. Relaxing this assumption will also lift the mass degeneracy within custodial multiplets, a welcome feature given the absence of clear experimental indications for (doubly-)charged states in the vicinity of the neutral states quoted in \cref{tab:significance}. Without specifying further the potential here, but assuming that it allows for a vacuum with a common VEV value for the two triplets, $\langle \chi^0 \rangle = \langle \xi^0 \rangle \sim u$ and two VEVs for the doublets, $\langle \phi_1^0 \rangle \sim v_1, \langle \phi_2^0 \rangle \sim v_2$, the physical mass eigenstates ${\cal H}_a$ will be related to the states of the gauge basis through an orthogonal transformation as follows:
\begin{equation}
\left(\begin{array}{c} {\cal H}_1 \\ {\cal H}_2 \\ {\cal H}_3 \\ {\cal H}_4 \end{array}\right) = {\cal X}_{4\times 4} 
\left(\begin{array}{c} {\rm Re} \phi_1^0 \\ {\rm Re} \phi_2^0 \\ {\rm Re} \chi^0 \\ \xi^0 \end{array}\right),\,  \ \  {\cal X}^\top {\cal X}= {\cal X} {\cal X}^\top =1, \ \ 
\left[{\cal X}\right]_{ij} \equiv x_{ij} . \label{eq:transition}
\end{equation}
Inverting \cref{eq:transition} and 
substituting in \cref{eq:Lkin} yields the ${\cal H}_a$'s couplings to $W^+W^-$ and $ZZ$:
\begin{equation}
v \kappa^{{\cal H}_{a}}_{W} = 
 v_1 x_{a 1} + v_2 x_{a 2} + 2 u (x_{a 3} + 
 \sqrt{2} x_{a 4}), \ v \kappa^{{\cal H}_{a}}_{Z} = v_1 x_{a 1} + v_2 x_{a 2} + 4 u x_{a 3}, \ a=1,...,4 . \label{eq:ggS}
\end{equation}
Here $v= (v_1^2+v_2^2+4u^2)^{1/2} \simeq \frac{246}{\sqrt{2}}$~GeV encodes the electroweak scale, and $\kappa^{{\cal H}_{a}}_{W}$ and $\kappa^{{\cal H}_{a}}_{Z}$ are the reduced couplings with respect to those of the SM. In the following we identify ${\cal H}_{a=1,...,4}$ with $\left( h_{95}, h_{125}, H_{320}, H_{650}\right)$. One should also bring in the couplings to fermions considering the four generic Yukawa configurations Type--I, II, X and Y. However, we will see that Type--II, X, and Y lead to the same conclusion since they have in common that at least one up or down quark or lepton couples differently from at least one other down or up fermion. When the Yukawa coupling of a fermion $f$ is carried by $\Phi_i$, its reduced coupling  to a scalar ${\cal H}_a$ is given by $\kappa^{{\cal H}_{a}}_f= \frac{v}{v_i} x_{ai}$ with no further freedom, being obtained by equating directly the SM expression of the fermion mass to that of the considered Yukawa Type. We can thus regroup the various Yukawa types as follows:
\begin{eqnarray}
& {\rm Type\!-\!II \, or \, X \, or \, Y}&: \ \   \kappa^{{\cal H}_{a}}_{f_d} = \frac{v}{v_1} x_{a1}\,, \ \ \kappa^{{\cal H}_{a}}_u= \frac{v}{v_2} x_{a2}\, \label{eq:typeIIXY} \\
& {\rm Type\!-\!I}&: \ \   \kappa^{{\cal H}_{a}}_{f_d} =  \kappa^{{\cal H}_{a}}_u= \frac{v}{v_2} x_{a2}\, \label{eq:typeI}
\end{eqnarray}
being understood that depending on the considered type, $f_d$ can stand for a down quark or charged lepton and $u$ for any up quark. Finally, one should also implement the constraint that
the matrix ${\cal X}$ in \cref{eq:transition} should be orthogonal, i.e. 
$ \sum_j x_{ij} x_{kj}  =\delta_{ik}   \label{eq:unitarity}$. 
\section{Possible Solutions}
\label{sec:pos-sol}
Satisfying simultaneously all the above equations while taking into account the experimental constraints of \cref{sec:indications} is quite involved. Note that \cref{eq:transition} presupposed no mixing with the fields imaginary parts---so as to avoid discussing possible sources of explicit CP-violation. The reduced couplings of the SM-like state 
$h_{125}$, satisfying  $\kappa^{h_{125}}_W \simeq \kappa^{h_{125}}_Z \simeq 1 $, lead to a strong contraint from \cref{eq:ggS} with $a=2$, namely the correlation $x_{2 3} \simeq \sqrt{2} x_{2 4}$. Then assuming Eq.~\ref{eq:typeIIXY} and $\kappa^{h_{125}}_f \simeq 1$ , plugged back in \cref{eq:ggS} and in the normality condition on the $x_{2i}$'s, one is lead to $u^2 \approx (1 - \kappa^{h_{125}}_Z) v^2$. Thus Yukawa types II, X and Y lead typically to too small values of $u$ that turn out to be inconsistent with real-valued $x_{ij}$ of the remaining states once the complete procedure is unfolded, in particular to fit large $\kappa^{H_{650}}_W$ as required by \cref{eq:WWH650min}. 
We stick hereafter to the Type-I Yukawa of Eq.~\ref{eq:typeI} and retain only real-valued VEVs so as not to deal with spontaneous CP-violation. Skimming over the full procedure that is somewhat technical we just give the main steps here: 

\noindent
1/ we choose 6 inputs: a/~$\kappa^{h_{125}}_W, \kappa^{h_{125}}_Z, \kappa^{h_{125}}_b$ (e.g. the central values of \cref{eq:CMS-SMkappas}), b/~$\kappa^{h_{95}}_W, \kappa^{h_{95}}_t$, scanning over these parameters in regions 	verifying simultaneously \cref{eq:CMSh95gg} and \cref{eq:CMSh95tautau}, c/~$\kappa^{H_{650}}_W$, a value in the domain given by \cref{eq:WWH650min}.

\noindent
2/ the input in a/ allows to determine uniquely $v_1,v_2$ and the $x_{2i}$'s using the equations discussed in this section, with $u$ chosen arbitrarily, however in a given domain of consistency. 

\noindent
3/ the knowledge of the 4-vector with components $x_{2i}$ constrains the 4-vector $x_{1i}$ to live on a unit sphere in the hyperplane orthogonal to $x_{2i}$. Combined with the input in b/ and \cref{eq:ggS} ($a=1$) and Eq.~\ref{eq:typeI}, the $x_{1i}$'s are fixed up to a possible discrete multiplicity.

\noindent
4/ the knowledge of the 4-vectors $x_{1i}$ and $x_{2i}$ constrains the 4-vector $x_{4i}$
to lie on a unit circle in the plane orthogonal to $x_{1i}$ and $x_{2i}$. Combined with the input in c/ and  the first of \cref{eq:ggS} (with $a=4$), fixes the $x_{4i}$'s up to a discrete multiplicity.

\noindent
5/ finally, from $x_{1i}$, $x_{2i}$ and $x_{4i}$ the 4-vector $x_{3i}$ is uniquely fixed up to a global sign, and all the remaining reduced couplings $\kappa^{h_{95}}_Z, \kappa^{H_{650}}_Z, \kappa^{H_{650}}_t, \kappa^{H_{320}}_W, \kappa^{H_{320}}_Z$ and  $\kappa^{H_{320}}_t,$ are predicted. 

\begin{figure}[h]
\begin{center}
{\includegraphics[width=.4\linewidth]{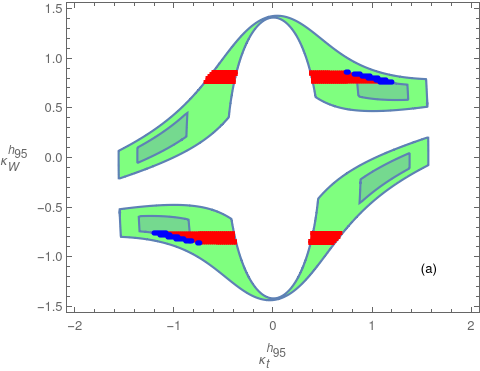}}
{\includegraphics[width=.4\linewidth]{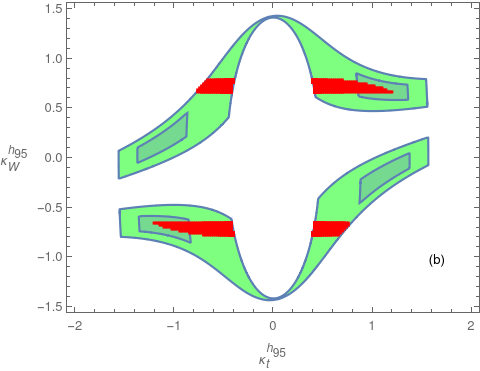}}
\end{center}
\caption{$\kappa_t^{h_{95}}$ versus 
$\kappa_W^{h_{95}}$: The light (dark)  green regions correspond to the 2(1)$\sigma$ constraints \cref{eq:CMSh95gg,eq:CMSh95tautau}; the red regions
correspond to no complex-valued couplings or mixings, and 
$\kappa_W^{H_{320}}, \kappa_Z^{H_{320}}$ satisfying \cref{eq:H320-constraints};
the blue regions correspond to overlaying the constraint \cref{eq:LEPh95} at the 2$\sigma$ level;
$u\simeq 78~{\rm GeV}, 
v_1 \simeq 16~{\rm GeV}, v_2 \simeq 76~{\rm GeV}$;
(a) $\kappa_W^{H_{650}} =.91$, 
(b) $\kappa_W^{H_{650}} =.97$
.}
\label{fig:kt95-kW95}
\end{figure}

\begin{figure}[h]
\begin{center}
{\includegraphics[width=.5\linewidth]{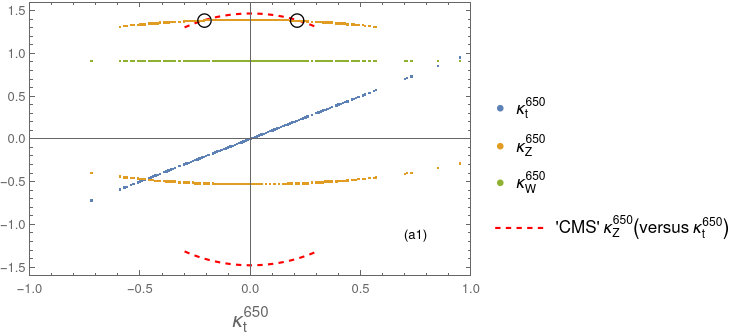}}
{\includegraphics[width=.33\linewidth]{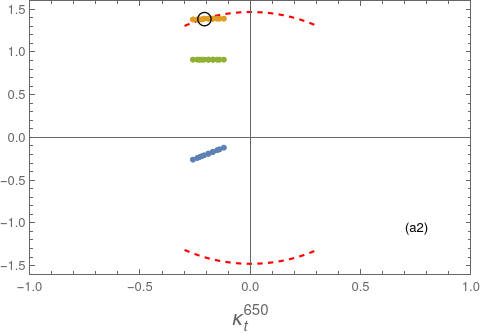}}
{\includegraphics[width=.33\linewidth]{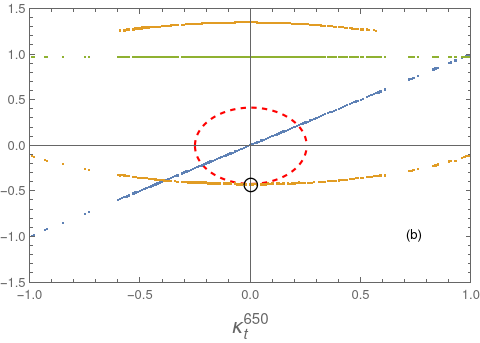}}
\end{center}
\caption{solutions for the $H_{650}$ reduced couplings: (a1), (a2) correspond to \cref{fig:kt95-kW95}(a); in (a2) \cref{eq:LEPh95} is imposed at the 2$\sigma$ level; the black circles indicate the points consistent with \cref{eq:xsec-data-SM} for $c=.78$; (b) corresponds to \cref{fig:kt95-kW95}(b).}
\label{fig:H650solutions}
\end{figure}

Following these five steps, we give here two examples of solutions satisfying `the constraints' and possibly `LEPh95'. 
Since we consider Type--I Yukawa, the couplings to t- and b-quark loops entering $h_{95} \to \gamma\gamma$ are the same as the coupling to $\tau$ entering $h_{95} \to \tau\tau$.   
\Cref{fig:kt95-kW95} shows the allowed domains satisfying  \cref{eq:CMSh95gg,eq:CMSh95tautau} and possibly `LEPh95', for the scanned over $\kappa_t^{h_{95}}$ and 
$\kappa_W^{h_{95}}$ , and fixing two values of $\kappa^{H_{650}}_W$; only the gauge boson and fermion virtual contributions to the $\gamma\gamma$ channel are included. Charged and doubly-charged states can also contribute, however in a model-dependent way controlled by the couplings in the potential, not specified here. They will however not change the typical pattern shown in \cref{fig:kt95-kW95}. \Cref{fig:H650solutions} then shows the pattern of the highly correlated values of $\kappa^{H_{650}}_Z$ and $\kappa^{H_{650}}_t$
for the given two input values of $\kappa^{H_{650}}_W$. Note the two-branch solution for 
$\kappa^{H_{650}}_Z$ in \cref{fig:H650solutions}(a1) \& (b), which is lifted
when the CMS indication for $H_{650}$  (\cref{eq:xsec-data-SM} and the red-dashed lines in \cref{fig:H650solutions}) is included. There is however an important sensitivity to $\kappa^{H_{650}}_W$; going from .91 to .97 changes drastically the solution values of $\kappa^{H_{650}}_Z$ while keeping $\kappa^{H_{650}}_t$ small enough to be consistent with the CMS scenario of `only-VBF' production. 
\begin{table}[h]
\caption{The ${\cal X}$ matrix and the reduced couplings for the 4 CP-even states; (A), (B) correspond to the black-circle solutions of \cref{fig:H650solutions}(a2),(b) respectively.}
\label{tab:solutions}
\begin{center}
\resizebox{.95\textwidth}{!}{
\begin{tabular}{cccccccc}
\multicolumn{8}{c}{(A)} \\
  \hline
$\fbox{c=.78}$ & $\phi_1^0$ & $\phi_2^0$ & $\chi^0$ & $\xi^0$ & $\kappa_t$ & $\kappa_Z$ & $\kappa_W$ \\ \hline \hline
 $h_{95}$ & $0.35$ & $-0.45$ & $0.36$ & $-0.74$ & $-1.04$ & $0.47$& $-0.78$ \\
 $h_{125}$ & $0.73$ & $0.43$ & $0.44$ & $0.29$ & $0.99$ & $1.04$ & $1.02$ \\
 $H_{320}$ & $0.26$& $-0.77$ & $-0.06$ & $0.57$&  $-1.78$ & $-0.42$& $0.36$ \\
 $H_{650}$ & $-0.52$ & $-0.09$ & $0.82$ & $0.21$ & $-0.21$ & $1.39$ & $0.91$ \\
 \hline
 \multicolumn{8}{c}{\fbox{$u\simeq 78~{\rm GeV}, v_1 \simeq 16~{\rm GeV}, v_2 \simeq 76~{\rm GeV}$}}
\end{tabular} \
\begin{tabular}{cccccccc}
\multicolumn{8}{c}{(B)} \\
  \hline
$\fbox{c=1}$& $\phi_1^0$ & $\phi_2^0$ & $\chi^0$ & $\xi^0$ & $\kappa_t$ & $\kappa_Z$ & $\kappa_W$ \\ \hline \hline
 $h_{95}$ & $0.42$& $0.24$& $-0.87$& $-0.1$& $0.56$& $-1.41$& $-0.76$ \\
 $h_{125}$ & $0.73$& $0.43$& $0.44$& $0.29$& $0.99$& $1.04$& $1.02$ \\
 $H_{320}$ & $-0.48$& $0.87$& $0.03$& $-0.12$& $1.99$& $0.38$& $0.21$ \\
 $H_{650}$ &  $-0.24$& $0.$& $-0.23$& $0.94$& $0.$& $-0.43$& $0.97$ \\
 \hline
 \multicolumn{8}{c}{\fbox{$u\simeq 78~{\rm GeV}, v_1 \simeq 16~{\rm GeV}, v_2 \simeq 76~{\rm GeV}$}}
\end{tabular}}
\end{center}
\end{table}
The outcome for the full CP-even sector is given in \cref{tab:solutions}.
While most of the input and predictions are consistent with the experimental indications, some are in tension: $\kappa^{H_{320}}_t$ is too large requiring even larger triple-scalar couplings to cope with the subdominance of the $t\bar t$ decay channel \cite{ATLAS:2022wti}; similarly, \cref{tab:solutions}(A) which is consistent with `LEPh95' shows a 
too large $\kappa^{H_{650}}_t$ while \cref{tab:solutions}(B) remains consistent, thus favoring the neglect of the `LEPh95' indication.
Finally, one should also inquire about possibe indications for charged ($H^\pm$) and doubly-charged states ($H^{\pm\pm}$). Searches for the latter decaying to same-sign $W$'s by ATLAS \cite{ATLAS:2023dbw} suggest an excess around 450~GeV, not seen though by CMS \cite{CMS:2021wlt} with slightly lower integrated luminosity. Interpreted within the GM, they both put stringent upper bounds on $u$, $\lsim 35$~GeV, which also applies here. The large $u$ found in our two examples, due to the requirement of real-valued parameters, can be lowered by increasing $\kappa^{H_{650}}_W$ to 1, but not sufficiently. The experimental limits can be relaxed in case of substancial decay fractions to $H^\pm W^\pm$ or 
$H^\pm H^\pm$ for sufficiently light singly charged Higgses. While some excesses for such
objects around 130 and 370~GeV have been reported we do not investigate them further in this note.

\section{Conclusions}
Should they stand the test of time, we have contemplated how some reported moderate excesses  at CMS and ATLAS, in particular the one around 650~GeV, would dictate a very specific extension of the SM with new neutral and (doubly-)charged scalars. Together with the recent CMS excess above $5\sigma$ \cite{CMS:2024ttbar} close to $A_{400}$, they might be signaling the dawn of a new LHC era. 

\section*{Acknowledgments}
We are indepted to Fran\c{c}ois Richard for having drawn our attention to the tentatively large significance of $H_{650}$ and the related need for doubly-charged scalars , as well as for many helpful discussions. 
AK is supported by the grant CRG/2023/000133 from ANRF, Govt. of India.
PM acknowledges funding from the SERB (Govt. of India) under grant SERB/CRG/2021/007579.
GM acknowledges partial support from the Marie Skłodowska-Curie grant agreement \newline 
No 101086085–ASYMMETRY. 
GM would like to thank ICISE and the PASCOS2024 
organizers for the invitation and for the great atmosphere during the conference and the social events; in particular the evening organized by the {\sl Union des Organisations d'Amitié Franco-Vietnamien de la province de Binh Dinh} where we toasted, sadly for the last time, with Prof PQ Hung.

\end{document}